\documentclass[a4paper,fleqn,usenatbib]{mnras}

\usepackage{newtxtext,newtxmath}

\usepackage[T1]{fontenc}
\usepackage{ae,aecompl}

\usepackage{graphicx,amsmath,color,amssymb}

\renewcommand{\vec}[1]{\boldsymbol{#1}}
\newcommand{\pop}{Phys.~Plasmas}
\newcommand{\jpp}{J.~Plasma Phys.}

\title[Kinetic plasma waves]{Dependence of kinetic plasma waves on ion-to-electron mass ratio and light-to-Alfv\'en speed ratio}

\author[D.~Verscharen et al.]{Daniel Verscharen,$^{1,2}$\thanks{E-mail: d.verscharen@ucl.ac.uk}
Tulasi N. Parashar,$^{3}$
S.~Peter Gary$^{4}$
\newauthor
and Kristopher G.~Klein$^{5}$
\\
$^{1}$Mullard Space Science Laboratory, University College London, Holmbury House, Holmbury St. Mary, Dorking RH5~6NT, UK\\
$^{2}$Space Science Center, University of New Hampshire, 8 College Road, Durham, NH 03824, USA\\
$^{3}$School of Chemical and Physical Sciences, Victoria University of Wellington, Gate 7, Kelburn Parade, Wellington 6012, \\ New Zealand\\
$^{4}$Space Science Institute, 4765 Walnut St, Boulder, CO 80301, USA\\
$^{5}$Lunar and Planetary Laboratory and Department of Planetary Sciences, University of Arizona, 1629 E University Blvd., \\ Tucson, AZ 85719, USA
}

\date{Accepted XXX. Received YYY; in original form ZZZ}

\pubyear{2020}

\begin{document}
\label{firstpage}
\pagerange{\pageref{firstpage}--\pageref{lastpage}}
\maketitle

\begin{abstract}
The magnetization $|\Omega_{\mathrm e}|/\omega_{\mathrm{e}}$ is an important parameter in plasma astrophysics, where $\Omega_{\mathrm e}$ and $\omega_{\mathrm{e}}$ are the electron gyro-frequency and electron plasma frequency, respectively. It only depends on the mass ratio $m_{\mathrm i}/m_{\mathrm e}$ and  the light-to-Alfv\'en speed ratio $c/v_{\mathrm{Ai}}$, where  $m_{\mathrm i}$
 ($m_{\mathrm e}$) is the ion (electron) mass, $c$ is the speed
 of light, and $v_{\mathrm{Ai}}$ is the ion Alfv\'en speed. Nonlinear numerical plasma models such as particle-in-cell simulations must often assume unrealistic
 values for $m_{\mathrm i}/m_{\mathrm e}$ and for
$c/v_{\mathrm{Ai}}$.  Because linear theory yields exact results for parametric scalings of wave properties at small amplitudes, we use linear theory to investigate the dispersion relations of  Alfv\'en/ion-cyclotron and fast-magnetosonic/whistler waves as prime examples for collective plasma behaviour depending on $m_{\mathrm i}/m_{\mathrm e}$ and $c/v_{\mathrm{Ai}}$. We analyse their dependence on $m_{\mathrm i}/m_{\mathrm e}$ and $c/v_{\mathrm{Ai}}$ in quasi-parallel and quasi-perpendicular
 directions of propagation with respect to the background magnetic field
 for a plasma with $\beta_j\sim1$, where $\beta_j$ is the ratio of the
 thermal to magnetic pressure for species $j$. Although their dispersion relations are largely
 independent of $c/v_{\mathrm{Ai}}$ for $c/v_{\mathrm{Ai}}\gtrsim 10$,
 the mass ratio $m_{\mathrm i}/m_{\mathrm e}$ has a strong effect at scales smaller than the ion inertial length. Moreover, we study the impact of relativistic electron effects on the dispersion relations. Based on our results, we recommend aiming for a more realistic value of $m_{\mathrm  i}/m_{\mathrm e}$ than for a more realistic value of $c/v_{\mathrm{Ai}}$
 in non-relativistic plasma simulations if such a choice is necessary, although relativistic and sub-Debye-length effects may require an additional adjustment of $c/v_{\mathrm{Ai}}$.
\end{abstract}


\section{Introduction}

The conditions and properties of plasmas in the universe vary greatly, ranging from
stellar cores to accretion discs, the intracluster medium, stellar winds, planetary magnetospheres, and laboratory
plasmas \citep{russell93,opher99,marsch06,uzdensky14,verscharen19,zhu19}. Some of these systems exhibit collisional relaxation timescales
comparable to or greater than the collective timescales of the plasma. The Vlasov--Maxwell set of equations describes 
such weakly collisional or collisionless plasmas adequately \citep{HasegawaSatoBook}. 
We write the Vlasov equation for species $j$ (e for electrons, i for ions) in dimensionless form as
\begin{equation}\label{vlasov}
\frac{\partial f_j}{\partial \tilde t}+\tilde{\vec v}\cdot \frac{\partial f_j}{\partial \tilde{\vec x}}+\frac{q_j}{q_{\mathrm i}}\frac{m_{\mathrm i}}{m_j}\left(\frac{c}{v_{\mathrm{Ai}}}\tilde{\vec E}+\tilde{\vec v}\times \tilde{\vec B}\right)\cdot \frac{\partial f_j}{\partial \tilde{\vec v}}=0,
\end{equation}
where $f_j$ is the distribution function of species $j$, $q_j$ and $m_j$ are the charge and mass of a particle of species $j$, $c$ is the speed of light, $\tilde t\equiv t\Omega_{\mathrm i}$ is the normalized time, $\tilde {\vec v}\equiv \vec v/v_{\mathrm{Ai}}$ is the normalized velocity, and $\tilde{\vec x}\equiv \vec x\Omega_{\mathrm i}/v_{\mathrm{Ai}}$ is the normalized spatial coordinate. Here, we define the signed gyro-frequency of species $j$ as 
\begin{equation}
\Omega_j\equiv \frac{q_jB_0}{m_jc}
\end{equation}
and the Alfv\'en speed of species $j$ as 
\begin{equation}\label{Alfven}
v_{\mathrm Aj}\equiv \frac{B_0}{\sqrt{4\pi n_jm_j}},
\end{equation}
where $B_0$ is the reference (background) magnetic field strength and $n_j$ is the density of species $j$.
In Eq.~(\ref{vlasov}), we normalize the electric and magnetic fields to $B_0$ as $\tilde {\vec E}\equiv \vec E/B_0$ and $\tilde{\vec B}\equiv \vec B/B_0$, respectively.

Eq.~(\ref{vlasov}) demonstrates the importance of the quantities $m_{\mathrm i}/m_j$ and $c/v_{\mathrm{Ai}}$ for the Lorentz-force term in Vlasov systems. Combining both quantities for a quasi-neutral two-species plasma with $n_{\mathrm i}\approx n_{\mathrm e}$, we define the magnetization of the plasma as the ratio
\begin{equation}
\frac{|\Omega_{\mathrm e}|}{\omega_{\mathrm e}}=\frac{v_{\mathrm{Ai}}}{c}\sqrt{\frac{m_{\mathrm i}}{m_{\mathrm e}}},
\end{equation}
where 
\begin{equation}
\omega_j\equiv \sqrt{\frac{4\pi n_jq_j^2}{m_j}}
\end{equation}
is the plasma frequency of species $j$. Kinetic simulations suggest that the ratio $|\Omega_{\mathrm e}|/\omega_{\mathrm{e}}$ is an
important parameter that controls the slope of observed particle spectra in, for example, pulsar wind nebulae, active galactic nuclei, and gamma-ray bursts  \citep{Cerutti13,melzaniAA14a,melzaniAA14b}.

Nonlinear computations such as particle-in-cell, Eulerian-Vlasov, and hybrid simulations have become standard tools to model astrophysical plasma
systems \citep{lapenta12,kunz14,markidis14,melzaniAA14a,germaschewski16,pezzi19}.
Computational constraints often prevent us from simulating realistic values for the multiple 
length scales and timescales involved in astrophysical plasma processes  \citep{vasconez14,franci15,ParasharApJ15,allanson19,cerri19,pecora19,pezzi19,verscharen19,matsukiyo20}. In the solar wind at 1~au \citep{MatthaeusJGR82,mccomas00,marsch06,klein19}, for
example, the energy-containing scale of the turbulence is of order $L\sim
10^6\,\mathrm{km} \sim 10^{11}\,\mathrm{cm}$, while the  Debye length
of species $j$, 
\begin{equation}\label{Debye}
\lambda_{j}\equiv \sqrt{\frac{k_{\mathrm B}T_{j}}{4\pi n_{j}q_{j}^2}},
\end{equation}
is of order $10^4\,\mathrm{cm}$ for both electrons and protons, where $k_{\mathrm B}$ is the Boltzmann
constant, and $T_{j}$ is the temperature of species $j$. This scale separation by seven orders of magnitude makes it difficult to simulate the system of size $L$ with a resolution of $\lambda_{j}$ based on 
present numerical capabilities.
Kinetic simulations with particle-in-cell and Eulerian Vlasov codes,
therefore, mostly focus on kinetic effects by resolving the relevant
spatial scales such as the $j$th species' inertial length 
\begin{equation}
 d_{j}\equiv \frac{c}{\omega_{j}}= \frac{v_{\mathrm{A}j}}{|\Omega_j|}
\end{equation}
or gyro-radius 
\begin{equation}
\rho_j \equiv \frac{w_{j}}{|\Omega_j|},
\end{equation}
 where 
 \begin{equation}\label{thermal_speed}
 w_{j} \equiv \sqrt{\frac{2k_{\mathrm B}T_j}{m_j}}
 \end{equation}
  is the thermal speed of species $j$.
Kinetic simulations often need to resolve all timescales from the
ion gyration ($\sim1/\Omega_{\mathrm i}$) up to the electron plasma
oscillation $(\sim 1/\omega_{\mathrm e})$.
%
%
%
%
For example, the ordering of kinetic length scales in the solar wind (assuming $\beta_j \sim 1$) is 
\begin{equation}
d_{\mathrm i} \sim \rho_{\mathrm i} \gg d_\mathrm{e} \sim \rho_\mathrm{e} \gg \lambda_\mathrm{e},
\end{equation}
where 
\begin{equation}
\beta_j\equiv \frac{8\pi n_j k_{\mathrm B}T_j}{B_0^2}
\end{equation}
is the plasma-$\beta$ (i.e., the ratio of thermal pressure to magnetic pressure) of species $j$.
Considering that 
\begin{equation}
\frac{d_{\mathrm i}}{d_{\mathrm e}}=\sqrt{\frac{m_{\mathrm i}}{m_{\mathrm e}}}
\end{equation}
 and 
 \begin{equation}
\frac{ d_{\mathrm e}}{\lambda_{\mathrm e}}=\sqrt{\beta_{\mathrm e}}\frac{\omega_{\mathrm e}}{|\Omega_{\mathrm e}|},
\end{equation}
achieving this ordering of scales in simulations requires sufficiently large values for $m_{\mathrm i}/m_{\mathrm e}$ and $c/v_{\mathrm{Ai}}$.
However, typical values of $|\Omega_{\mathrm e}|/\omega_{\mathrm{e}}$ used in fully 
kinetic studies of the solar wind are $|\Omega_{\mathrm e}|/\omega_{\mathrm{e}}\sim 0.1-1$ 
 \citep{KarimabadiPP13, saito14,ParasharApJ15, ParasharJPP15,groselj18,parashar19,roytershteyn19}, in contrast to the 
realistic values $|\Omega_{\mathrm e}|/\omega_{\mathrm{e}} \sim 10^{-3}-10^{-2}$ \citep[for example, see][table~1]{verscharen19}. Therefore, it is important to understand and quantify the effects of artificially large values of $|\Omega_{\mathrm e}|/\omega_{\mathrm{e}}$ on the  dynamics of the system. We study this particular question in the context of linear plasma waves as some of the most fundamental building blocks of plasma dynamics. Linear wave theory has the unique advantage that it yields exact numerical results under the assumption that the fluctuation amplitude is small, whereas nonlinear computations such as particle-in-cell simulations yield only approximate scaling relations. Therefore, linear theory provides a standard of comparison against which small-amplitude simulations can be tested. The goal of our work is to provide ground truth for such comparisons between nonlinear computations and linear theory, and to raise awareness for the inaccuracies from artificially decreasing $m_{\mathrm i}/m_{\mathrm e}$ and $c/v_{\mathrm{Ai}}$ in nonlinear computations.




\section{Plasma waves in linear theory}

We use the numerical code \textsc{nhds} \citep{verscharen18b} to solve the kinetic dispersion relation in a quasi-neutral two-species plasma consisting of isotropic and non-drifting Maxwellian ions and electrons. The Maxwellian distribution for species $j$ is given by
\begin{equation}\label{maxwell}
f_{0j}=\frac{n_j}{\pi^{3/2}w_j^3}\exp\left(-\frac{v^2}{w_j^2}\right),
\end{equation}
where $v$ is the velocity coordinate and $w_j$ is the thermal speed as defined in Eq.~(\ref{thermal_speed}).
 \textsc{nhds}, like other solvers of the hot-plasma dispersion relation  \citep{roennmark82,gary93,klein15}, determines the non-trivial solutions to the wave equation,
\begin{equation}\label{waveeq}
\vec n\times \left(\vec n\times \vec E\right)+\vec \varepsilon\vec E=0,
\end{equation}
where $\vec n\equiv \vec kc/\omega$, $\vec E$ is the Fourier transform of the electric field, $\vec k$ is the wavevector,  $\omega$ is the (complex) wave frequency,
\begin{equation}\label{epsilon}
\vec \varepsilon\equiv \vec {1}+\sum \limits_j\vec \chi_j
\end{equation}
is the dielectric tensor, and $\vec \chi_j$ is the contribution of species $j$ to the susceptibility  \citep[for details, see][]{stix92}. \textsc{nhds} calculates the tensor $\vec \chi_j$ from the linearised Vlasov equation [see Eq.~(\ref{vlasov})], Maxwell's equations, and Eq.~(\ref{maxwell}).

We set $\beta_{\mathrm i}=\beta_{\mathrm e}=1$. We investigate the Alfv\'en/ion-cyclotron (A/IC) wave and the fast-magnetosonic/whistler (FM/W) wave in both quasi-parallel ($\theta = 0.001^{\circ}$) and quasi-perpendicular ($\theta=89^{\circ}$) directions of propagation with respect to the background magnetic field $\vec B_0$, where $\theta$ is the angle between $\vec k$ and $\vec B_0$. These waves are important normal modes in plasmas  \citep{ofman10,marsch11,verscharen12,boldyrev13,yoon15,comisel16,cerri17,wu19,zhu19}. We show solutions for the real part $\omega_{\mathrm r}$ of the wave frequency  and for its imaginary part  $\gamma$ as functions of the wavenumber $k$. We only plot solutions when $|\gamma|\le \omega_{\mathrm r}$. We normalize all length scales in units of $d_{\mathrm i}$ and all frequencies in units of $\Omega_{\mathrm i}$. In addition, we indicate the values of $kd_{\mathrm e}=1$ for the different mass ratios in units of $kd_{\mathrm i}$.

\subsection{Dependence on ion-to-electron mass ratio}

\begin{figure}
 \includegraphics[width=\columnwidth]{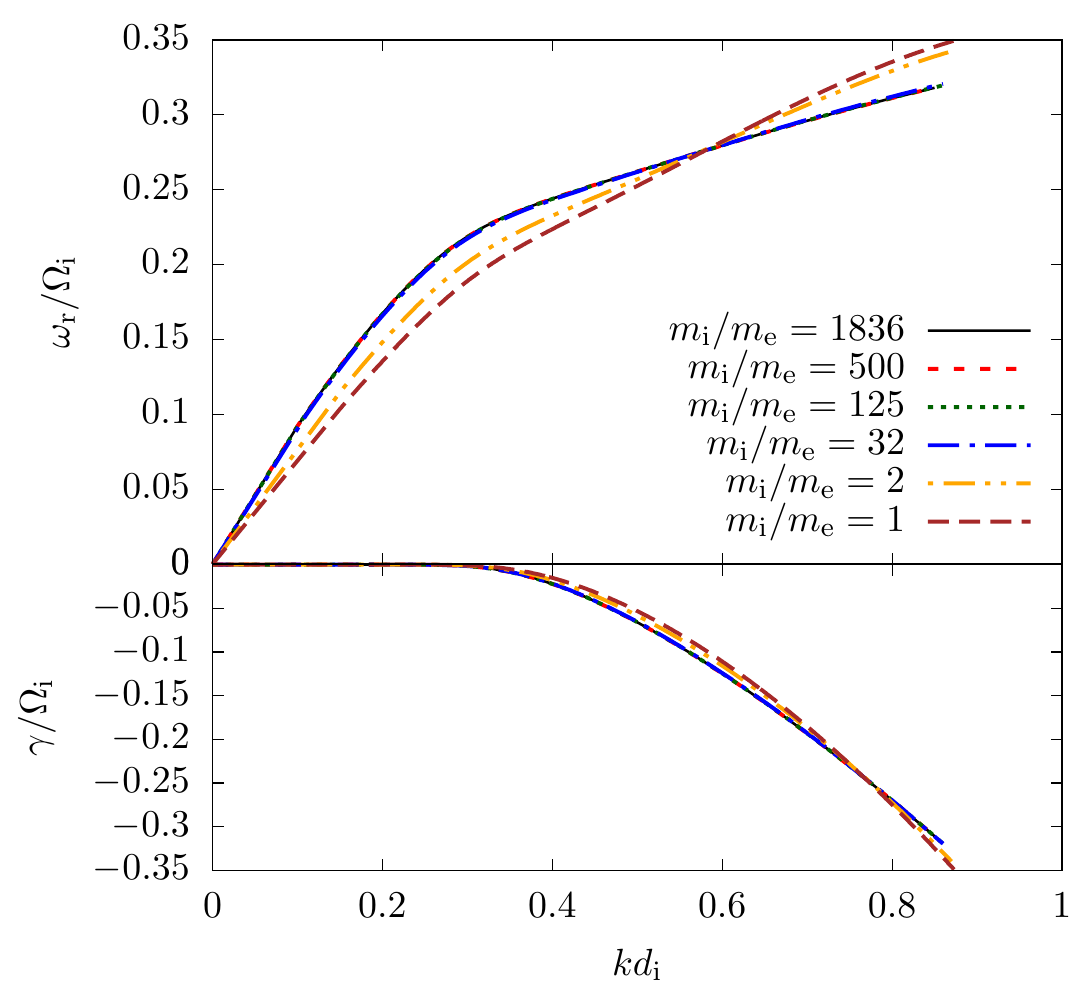}%
 \caption{Dispersion relations of the A/IC wave in quasi-parallel propagation for different values of $m_{\mathrm i}/m_{\mathrm e}$. The light-to-Alfv\'en speed ratio is fixed at $c/v_{\mathrm{Ai}}=10^4$. The top panel shows the normalized real part of the wave frequency, and the bottom panel shows the normalized imaginary part of the wave frequency as functions of the normalized wavenumber $k$. None of the solutions reaches $kd_{\mathrm e}=1$ while $|\gamma|/\omega_{\mathrm r}<1$. \label{parallel_AIC_mass}}%
\end{figure}

We first investigate the dependence of the A/IC and FM/W dispersion relations on $m_{\mathrm i}/m_{\mathrm e}$. We fix the light-to-Alfv\'en speed ratio at $c/v_{\mathrm{Ai}}=10^4$ and vary $m_{\mathrm i}/m_{\mathrm e}$ between 1836 (i.e., the realistic value for the proton-to-electron mass ratio, $m_{\mathrm p}/m_{\mathrm e}$) and 1 (i.e., the value for an electron--positron plasma). The magnetization, therefore, varies from $|\Omega_{\mathrm e}|/\omega_{\mathrm{e}}=10^{-4}$ to $|\Omega_{\mathrm e}|/\omega_{\mathrm{e}}\approx4\times 10^{-3}$. In Fig.~\ref{parallel_AIC_mass}, we show the dispersion relations for the A/IC wave in quasi-parallel propagation for different values of $m_{\mathrm i}/m_{\mathrm e}$.  Both the real part of the normalized frequency and its imaginary part do not vary significantly with $m_{\mathrm i}/m_{\mathrm e}$ since the dispersion relation of the quasi-parallel A/IC wave is dominated by ion dynamics (unless $kd_{\mathrm i}\ll 1$) and by the onset of ion-cyclotron damping. The cyclotron-resonance condition for particles of species $j$ is given by
\begin{equation}\label{cyclres}
\omega_{\mathrm r}=k_{\parallel}v_{\parallel}+n\Omega_{j},
\end{equation}
where $v_{\parallel}$ is the speed of the resonant particles in the $\vec B_0$ direction, and $n$ is the integer order of the cyclotron resonance. Cyclotron-resonant interactions are typically most efficient for $n=\pm 1$. Since the quasi-parallel A/IC wave is left-hand polarized, electrons cannot cyclotron-resonate with this wave through the otherwise most efficient $n=-1$ resonance. We note that the resonant cut-off of the dispersion relation for $\omega_{\mathrm r}\rightarrow \Omega_{\mathrm i}$ is also present in the cold-plasma limit \citep{stix92}. 

\begin{figure}
 \includegraphics[width=\columnwidth]{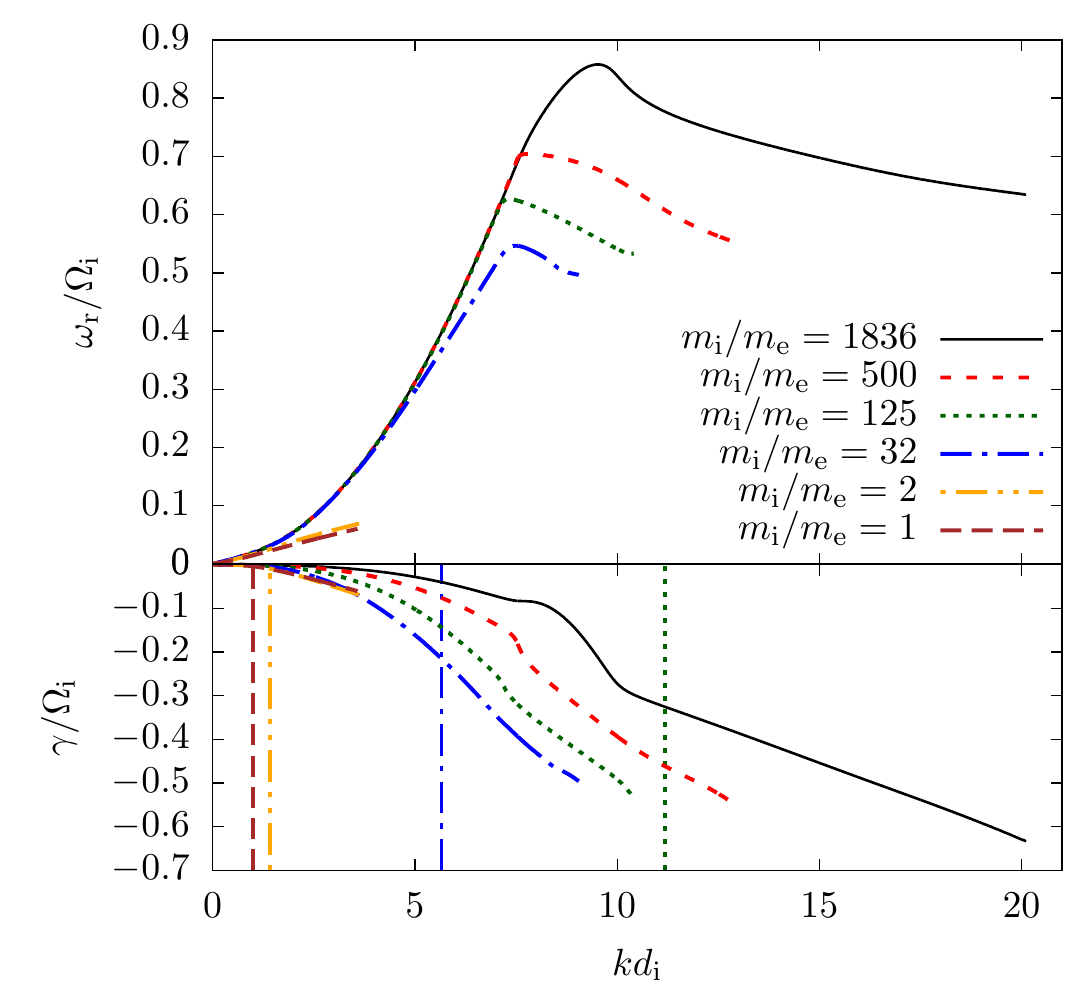}%
 \caption{Dispersion relations of the A/IC wave in quasi-perpendicular propagation for different values of $m_{\mathrm i}/m_{\mathrm e}$. The light-to-Alfv\'en speed ratio is fixed at $c/v_{\mathrm{Ai}}=10^4$. The top panel shows the normalized real part of the wave frequency, and the bottom panel shows the normalized imaginary part of the wave frequency as functions of the normalized wavenumber $k$. The vertical lines in the bottom panel mark the positions of $kd_{\mathrm e}=1$ for the given mass ratios. \label{perp_AIC_mass}}%
\end{figure}

In Fig.~\ref{perp_AIC_mass}, we show the dispersion relations for the A/IC wave in quasi-perpendicular direction of propagation. This mode shows a characteristic maximum in $\omega_{\mathrm r}$. Both the normalized frequency and the normalized wavenumber associated with this maximum decrease with decreasing $m_{\mathrm i}/m_{\mathrm e}$. On the other hand, the magnitude of the damping rate $|\gamma|/\Omega_{\mathrm i}$ increases with decreasing $m_{\mathrm i}/m_{\mathrm e}$. With decreasing $m_{\mathrm i}/m_{\mathrm e}$, the electron thermal speed decreases, leading to a non-isothermal electron behaviour and an increase in the electron inertia. This effect becomes important for the quasi-perpendicular A/IC wave since it is a compressive mode at $k\rho_{\mathrm i}\gtrsim 1$  \citep{schekochihin09,hunana13,chen17}. However, the behaviour of the quasi-perpendicular A/IC wave does not vary significantly over the explored $m_{\mathrm i}/m_{\mathrm e}$ range at small $kd_{\mathrm i}\lesssim 1$. For $m_{\mathrm i}/m_{\mathrm e}\lesssim 100$, the quasi-perpendicular A/IC mode damps heavily (i.e., $|\gamma|\gtrsim \omega_{\mathrm r}$) even before reaching electron scales at $kd_{\mathrm e}\ge 1$. For larger values of $m_{\mathrm i}/m_{\mathrm e}$, it extends without significant damping beyond $kd_{\mathrm e}= 1$.

\begin{figure}
 \includegraphics[width=\columnwidth]{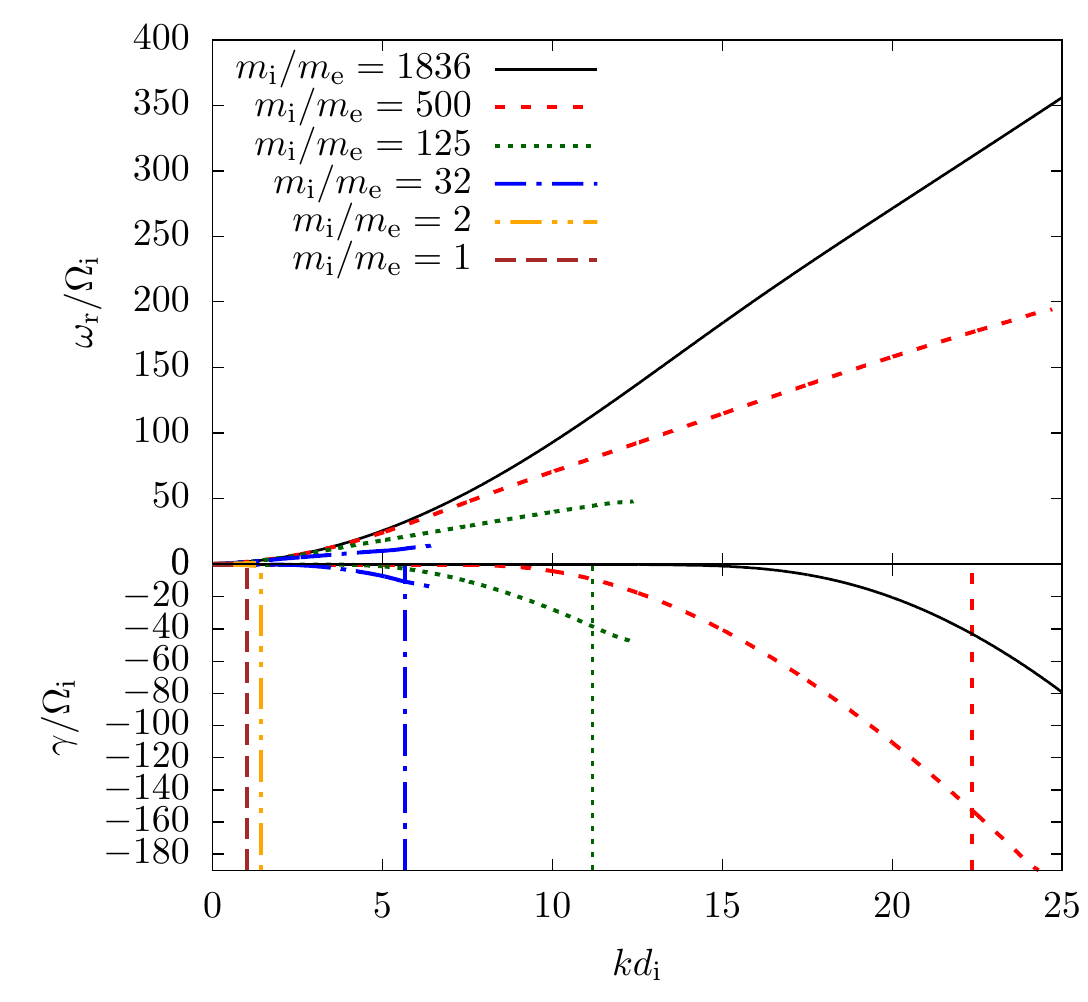}%
 \caption{Dispersion relations of the FM/W wave in quasi-parallel propagation for different values of $m_{\mathrm i}/m_{\mathrm e}$. The light-to-Alfv\'en speed ratio is fixed at $c/v_{\mathrm{Ai}}=10^4$. 
The plot follows the same format as Fig.~\ref{perp_AIC_mass}. \label{parallel_FMW_mass}}%
\end{figure}

In Fig.~\ref{parallel_FMW_mass}, we show the dispersion relations for the FM/W wave in quasi-parallel direction of propagation. This mode shows the typical quadratic whistler-mode behaviour at $1\lesssim kd_{\mathrm i}\lesssim \sqrt{m_{\mathrm i}/m_{\mathrm e}}$. The FM/W wave is dominated by electron dynamics unless $kd_{\mathrm i}\lesssim 1$. With decreasing $m_{\mathrm i}/m_{\mathrm e}$, $\omega_{\mathrm r}/\Omega_{\mathrm i}$ decreases while $|\gamma|/\Omega_{\mathrm i}$ increases. 
We attribute this behaviour to increasing cyclotron damping by electrons, which is, for the quasi-parallel FM/W wave, most efficient when $n=-1$ in Eq.~(\ref{cyclres}).  
Thermal electrons with $v_{\parallel}\approx -w_{\mathrm e}$ thus efficiently resonate with the quasi-parallel FM/W mode when
\begin{equation}\label{cycloelec}
\left(\frac{\omega_{\mathrm r}}{\Omega_{\mathrm i}}\right)\approx -\left(kd_{\mathrm i}\right)\cos\theta\sqrt{\frac{m_{\mathrm i}}{m_{\mathrm e}}\beta_{\mathrm e}}+\frac{m_{\mathrm i}}{m_{\mathrm e}}
\end{equation}
in our normalization. In the electron--positron case with $m_{\mathrm i}/m_{\mathrm e}=1$, a significant number of electrons resonate with the FM/W mode at $\omega_{\mathrm r}/\Omega_{\mathrm i}=\omega_{\mathrm r}/|\Omega_{\mathrm e}|\lesssim 1$ and at $kd_{\mathrm i}=kd_{\mathrm e}\lesssim 1$ \citep[see fig.~2 by][]{gary09}. Like in the case of the quasi-parallel A/IC wave, the resonant cut-off of the dispersion relation for $\omega_{\mathrm r}\rightarrow |\Omega_{\mathrm e}|$ is also present in the cold-plasma limit \citep{stix92}. The quasi-parallel FM/W mode reaches the wavenumber range $kd_{\mathrm e}\ge 1$ with $|\gamma|<\omega_{\mathrm r}$ for all mass ratios shown in Fig.~\ref{parallel_FMW_mass}.

\begin{figure}
 \includegraphics[width=\columnwidth]{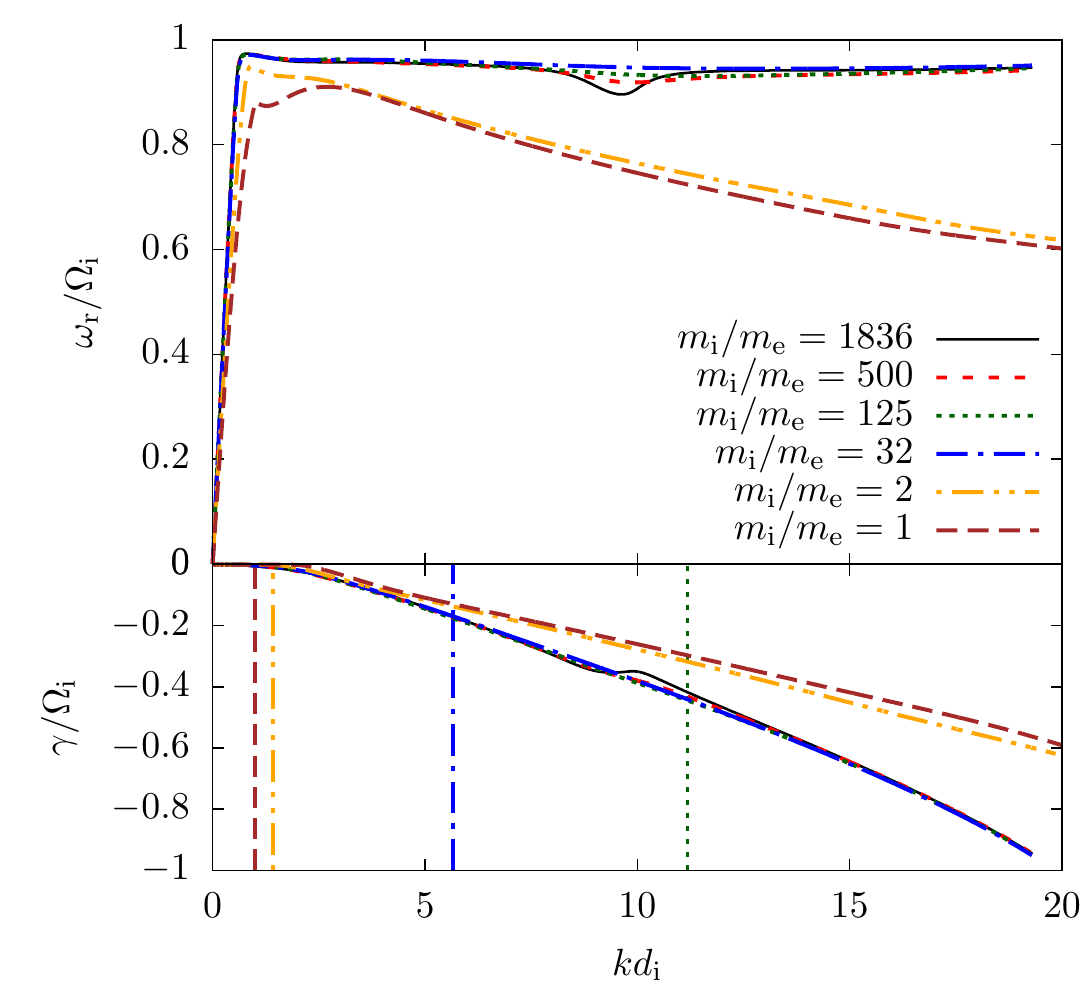}%
 \caption{Dispersion relations of the FM/W wave in quasi-perpendicular propagation for different values of $m_{\mathrm i}/m_{\mathrm e}$. The light-to-Alfv\'en speed ratio is fixed at $c/v_{\mathrm{Ai}}=10^4$. 
The plot follows the same format as Fig.~\ref{perp_AIC_mass}. \label{perp_FMW_mass}}%
\end{figure}

In Fig.~\ref{perp_FMW_mass}, we show the dispersion relations for the FM/W wave in quasi-perpendicular direction of propagation. The dispersion relation does not depend significantly on $m_{\mathrm i}/m_{\mathrm e}$ for $m_{\mathrm i}/m_{\mathrm e}\gtrsim 10$. The quasi-perpendicular FM/W wave reaches the ion-Bernstein regime at $\omega_{\mathrm r}/ \Omega_{\mathrm i} \approx 1$. In the perpendicular limit ($\theta\rightarrow 90^{\circ}$), the susceptibilities $\vec{\chi}_j$ for both ions and electrons include sums over all $n$ of contributions that are each proportional to  $1/\left(\omega_{\mathrm r}-n\Omega_j\right)$, leading to this resonant behaviour. As long as $m_{\mathrm i}/m_{\mathrm e}\gg 1$, electron harmonics do not interfere with the ion harmonics since the electrons' contributions to $\vec \varepsilon$ decrease with increasing $n$. For $m_{\mathrm i}/m_{\mathrm e}=2$, however, the second-order ion resonance makes a contribution to $\vec \varepsilon$ that is comparable with the first-order electron resonance. The quasi-perpendicular FM/W mode reaches the wavenumber range $kd_{\mathrm e}\ge 1$ with $|\gamma| < \omega_{\mathrm r}$ for all shown parameter combinations with $m_{\mathrm i}/m_{\mathrm e}\le 125$.

\subsection{Dependence on light-to-Alfv\'en speed ratio}

We now investigate the dependence of the dispersion relation on $c/v_{\mathrm {Ai}}$. For this study, we fix the mass ratio at $m_{\mathrm i}/m_{\mathrm e}=100$ and $m_{\mathrm i}/m_{\mathrm e}=1$ (electron--positron plasma). We then vary $c/v_{\mathrm{Ai}}$ between $10^4$ and $3$ for $m_{\mathrm i}/m_{\mathrm e}=100$ and between $10$ and $3$ for $m_{\mathrm i}/m_{\mathrm e}=1$. The magnetization, therefore, varies from $|\Omega_{\mathrm e}|/\omega_{\mathrm{e}}=10^{-3}$ to $|\Omega_{\mathrm e}|/\omega_{\mathrm{e}}\approx 3.33$ for $m_{\mathrm i}/m_{\mathrm e}=100$ and from $|\Omega_{\mathrm e}|/\omega_{\mathrm{e}}=0.1$ to $|\Omega_{\mathrm e}|/\omega_{\mathrm{e}}\approx 3.33$ for $m_{\mathrm i}/m_{\mathrm e}=1$.
Applying our normalization to Eq.~(\ref{waveeq}) and multiplying with $v_{\mathrm{Ai}}^3/c^3$ removes the $c/v_{\mathrm{Ai}}$ dependence from all terms in Eq.~(\ref{waveeq}) except the unit tensor in the definition of $\vec \varepsilon$ in Eq.~(\ref{epsilon}). This unit tensor represents the displacement current in Amp\`ere's law, and thus $c/v_{\mathrm{Ai}}$ controls the effects due to the displacement current in our normalization.

Relativistic plasma effects become important if
\begin{equation}\label{relgen}
\frac{k_{\mathrm B}T_j}{m_jc^2}\gtrsim 1.
\end{equation}
In our normalization,
\begin{equation}
\frac{k_{\mathrm B}T_j}{m_jc^2}=\frac{1}{2}\beta_j\frac{m_{\mathrm i}}{m_j}\left(\frac{v_{\mathrm{Ai}}}{c}\right)^2.
\end{equation}
For $\beta_j=1$, relativistic effects, therefore, become important if
\begin{equation}\label{relcond}
\frac{c}{v_{\mathrm{Ai}}}\lesssim \sqrt{\frac{m_{\mathrm i}}{m_j}}.
\end{equation}
We note that relativistic effects already modify the plasma behaviour if $k_{\mathrm B}T_j/m_jc^2\gtrsim 0.1$. For the cases under consideration, ions do not fulfil Eq.~(\ref{relcond})\footnote{According to Eq.~(\ref{Alfven}), it is possible that $v_{\mathrm{Ai}}/c>1$. In this case, all plasma waves have subluminal group velocities $<v_{\mathrm{Ai}}$, so that $v_{\mathrm{Ai}}$ loses its meaning as the group velocity of any wave mode.}. Electrons, however, easily satisfy Eq.~(\ref{relcond}) even when using unrealistic mass ratios in plasma models. We, therefore, calculate the true relativistic dispersion relation with the \textsc{alps} code \citep{verscharen18} for comparison with the non-relativistic solutions from \textsc{nhds}. For our \textsc{alps} calculations, we assume that the plasma consists of an isotropic, non-relativistic ion species with a Maxwellian distribution function  [see Eq.~(\ref{maxwell})] and an isotropic, relativistic electron species with a J\"uttner distribution \citep{juttner11},
\begin{equation}
f_{0j}=\frac{n_j}{2\pi m_j^3cw_j^2 K_2(2c^2/w_j^2)}\exp\left(-2\frac{c^2}{w_j^2}  \frac{1}{\sqrt{1-v^2/c^2}}\right),
\end{equation}
where $K_2$ is the modified Bessel function of the second kind.
We use the following parameters in our \textsc{alps} calculations \citep[for definitions, see][]{verscharen18}: $c/v_{\mathrm {Ai}}=10$, $\beta_{\mathrm i}=\beta_{\mathrm e}=1$, $m_{\mathrm i}/m_{\mathrm e}=100$, $N_{\perp}=100$, $N_{\parallel}=200$, $N_{\Gamma}=100$, $N_{\bar p_{\parallel}}=500$, $P_{\max,\parallel \mathrm i}=P_{\max,\perp \mathrm i}=10m_{\mathrm i}v_{\mathrm {Ai}}$, $P_{\max,\parallel \mathrm e}=P_{\max,\perp\mathrm e}=0.6m_{\mathrm i}v_{\mathrm {Ai}}$,  $J_{\max}=10^{-45}$, $M_{\mathrm I}=3$, $M_{\mathrm P}=100$, and $t_{\mathrm{lim}}=0.01$. With these parameters,
\begin{equation}
\frac{k_{\mathrm B}T_{\mathrm e}}{m_{\mathrm e}c^2}=0.5,
\end{equation}
corresponding to a mildly relativistic regime for the electrons, while the ions are non-relativistic.

\begin{figure}
 \includegraphics[width=\columnwidth]{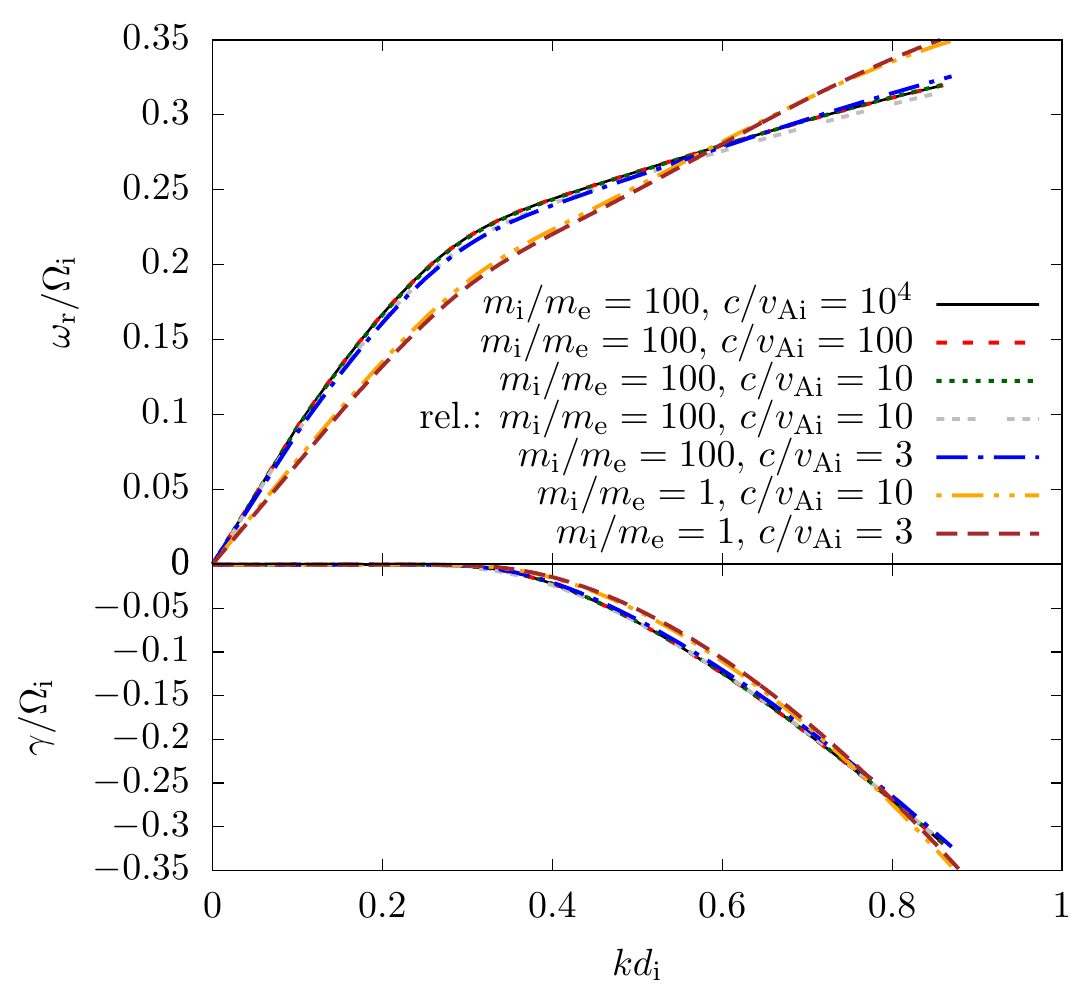}%
 \caption{Dispersion relations of the A/IC wave in quasi-parallel propagation for different values of $c/v_{\mathrm{Ai}}$. We fix the ion-to-electron mass ratio at $m_{\mathrm i}/m_{\mathrm e}=100$ and $m_{\mathrm i}/m_{\mathrm e}=1$. The top panel shows the normalized real part of the wave frequency, and the bottom panel shows the normalized imaginary part of the wave frequency as functions of the normalized wavenumber $k$. Only the solution marked as `rel.' includes relativistic effects. \label{parallel_AIC_vA}}%
\end{figure}

In Fig.~\ref{parallel_AIC_vA}, we show the dispersion relations for the A/IC wave in quasi-parallel propagation. We see that this mode's behaviour is largely independent of $c/v_{\mathrm{Ai}}$. Even for $c/v_{\mathrm{Ai}}=3$, its deviation from the high-$c/v_{\mathrm{Ai}}$ case is insignificant. We note that the A/IC wave is strongly damped at scales $kd_{\mathrm i}\gtrsim 1$, so that potential modifications at smaller scales due to smaller $c/v_{\mathrm{Ai}}$ are not relevant in the propagating regime of the A/IC wave. In the quasi-parallel A/IC wave with $m_{\mathrm i}\gg m_{\mathrm e}$, the ions carry most of the polarization current unless $kd_{\mathrm i}\ll 1$, in which case both ions and electrons contribute almost equally to the polarization current. Therefore, our relativistic calculation for $m_{\mathrm i}/m_{\mathrm e}=100$ and $c/v_{\mathrm{Ai}}=10$ is almost identical to the non-relativistic case.

\begin{figure}
 \includegraphics[width=\columnwidth]{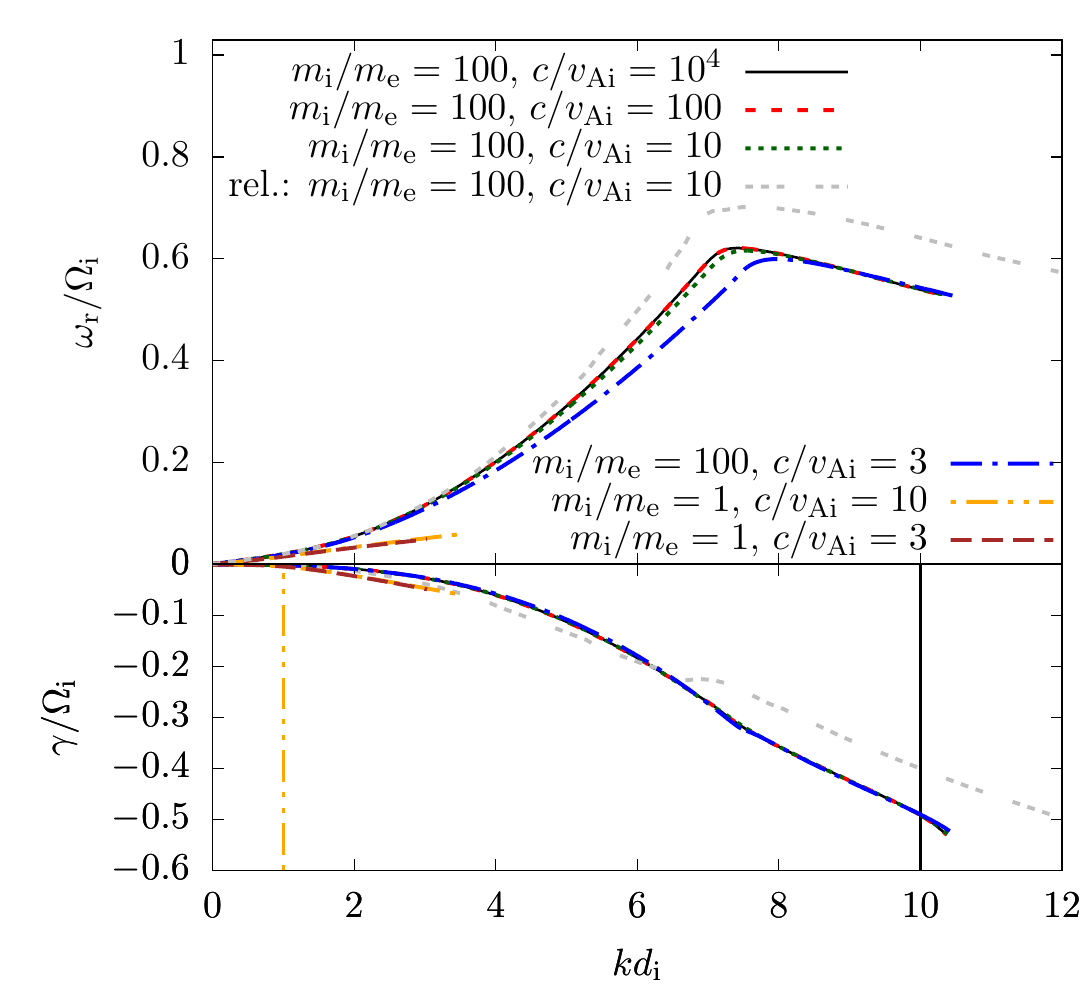}%
 \caption{Dispersion relations of the A/IC wave in quasi-perpendicular propagation for different values of $c/v_{\mathrm{Ai}}$. We fix the ion-to-electron mass ratio at $m_{\mathrm i}/m_{\mathrm e}=100$ and $m_{\mathrm i}/m_{\mathrm e}=1$. The top panel shows the normalized real part of the wave frequency, and the bottom panel shows the normalized imaginary part of the wave frequency as functions of the normalized wavenumber $k$. In the bottom panel, the vertical black line marks the position of $kd_{\mathrm e}=1$ for $m_{\mathrm i}/m_{\mathrm e}=100$, and the vertical orange line marks the position of $kd_{\mathrm e}=1$ for $m_{\mathrm i}/m_{\mathrm e}=1$.  Only the solution marked as `rel.' includes relativistic effects. \label{perp_AIC_vA}}%
\end{figure}

In Fig.~\ref{perp_AIC_vA}, we show the dispersion relations for the A/IC wave in quasi-perpendicular propagation. Like in the quasi-parallel case, the $c/v_{\mathrm{Ai}}$ dependences of $\omega_{\mathrm r}/\Omega_{\mathrm i}$ and $\gamma/\Omega_{\mathrm i}$ are insignificant. We only find a small deviation of $\omega_{\mathrm r}$ for $c/v_{\mathrm{Ai}}=3$ at $kd_{\mathrm i}\gtrsim 4$. Figs.~\ref{parallel_AIC_vA} and \ref{perp_AIC_vA} suggest that the behaviour of the A/IC wave is largely independent of $c/v_{\mathrm{Ai}}$ in the non-relativistic case. However, the introduction of relativistic electrons changes the dispersion relation of the quasi-perpendicular A/IC wave at $kd_{\mathrm i}\gtrsim 6$, leading to an increase in $\omega_{\mathrm r}/\Omega_{\mathrm i}$ and a decrease in $|\gamma|/\Omega_{\mathrm i}$ compared to the non-relativistic case. We associate this behaviour with relativistic modifications to the electron-pressure-gradient force, which plays an important role in quasi-perpendicular A/IC waves at $k\rho_{\mathrm i}\gtrsim 1$.

\begin{figure}
 \includegraphics[width=\columnwidth]{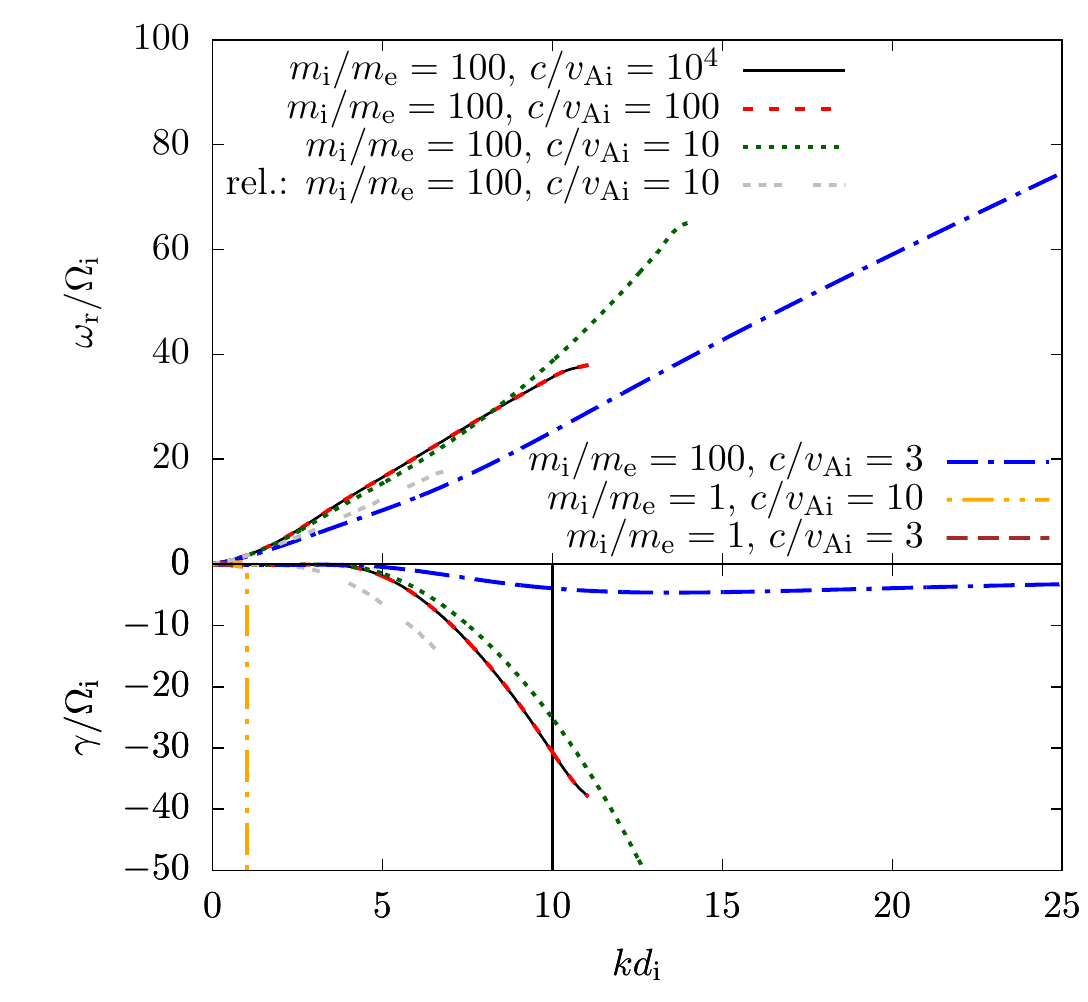}%
 \caption{Dispersion relations of the FM/W wave in quasi-parallel propagation for different values of $c/v_{\mathrm{Ai}}$. We fix the ion-to-electron mass ratio at $m_{\mathrm i}/m_{\mathrm e}=100$ and $m_{\mathrm i}/m_{\mathrm e}=1$. The plot follows the same format as Fig.~\ref{perp_AIC_vA}. \label{parallel_FMW_vA}}%
\end{figure}

In Fig.~\ref{parallel_FMW_vA}, we show the dispersion relations for the FM/W wave in quasi-parallel propagation. For $c/v_{\mathrm{Ai}}\ge 10$, the normalized dispersion relation is largely independent of $c/v_{\mathrm{Ai}}$. For $c/v_{\mathrm{Ai}}=3$, however, the behaviour strongly deviates. We especially find that $|\gamma|/\Omega_{\mathrm i}$ is significantly smaller than in cases with larger $c/v_{\mathrm{Ai}}$. Although we clearly identify this mode as purely right-handed in polarization, it is not affected by electron-cyclotron resonant damping in the same way as the other examples. Even the FM/W-wave-typical asymptotic high-frequency behaviour at $\omega_{\mathrm r}\approx |\Omega_{\mathrm e}|=100\Omega_{\mathrm i}$ is not present.  Our relativistic solution for $m_{\mathrm i}/m_{\mathrm e}=100$ and $c/v_{\mathrm{Ai}}=10$ also deviates from our non-relativistic solution with the same parameters. Since electrons carry most of the polarization current in the quasi-parallel FM/W wave for $kd_{\mathrm i}\gtrsim 1$, the introduction of relativistic effects leads to a slight increase in $\omega_{\mathrm r}/\Omega_{\mathrm i}$ and a more significant increase in  $|\gamma|/\Omega_{\mathrm i}$ compared to the non-relativistic case. In the relativistic case, the cyclotron-resonance condition from Eq.~(\ref{cyclres}) is modified by the relativistic mass-dependence of $\Omega_j$:
\begin{equation}\label{rescondrel}
\omega_{\mathrm r}=k_{\parallel}v_{\parallel}+n\Omega_{j}\sqrt{1-\frac{v^2}{c^2}}.
\end{equation}
For a  wave with fixed $\omega_{\mathrm r}$ and $k_{\parallel}$, Eq.~(\ref{rescondrel}) suggests that  $v_{\parallel}$ is not equal for all resonant particles, as it is the case in Eq.~(\ref{cyclres}). Instead, the resonance condition in Eq.~(\ref{rescondrel}) is fulfilled by particles with a range of $v_{\parallel}$, and the resonance condition now also depends on the particles' $v_{\perp}$ through $v^2=v_{\perp}^2+v_{\parallel}^2$. In the relativistic case  of the quasi-parallel FM/W wave, a larger number of electrons fulfil the resonance condition with $n=-1$ than in the non-relativistic case, so that $|\gamma|/\Omega_{\mathrm i}$ increases.

\begin{figure}
 \includegraphics[width=\columnwidth]{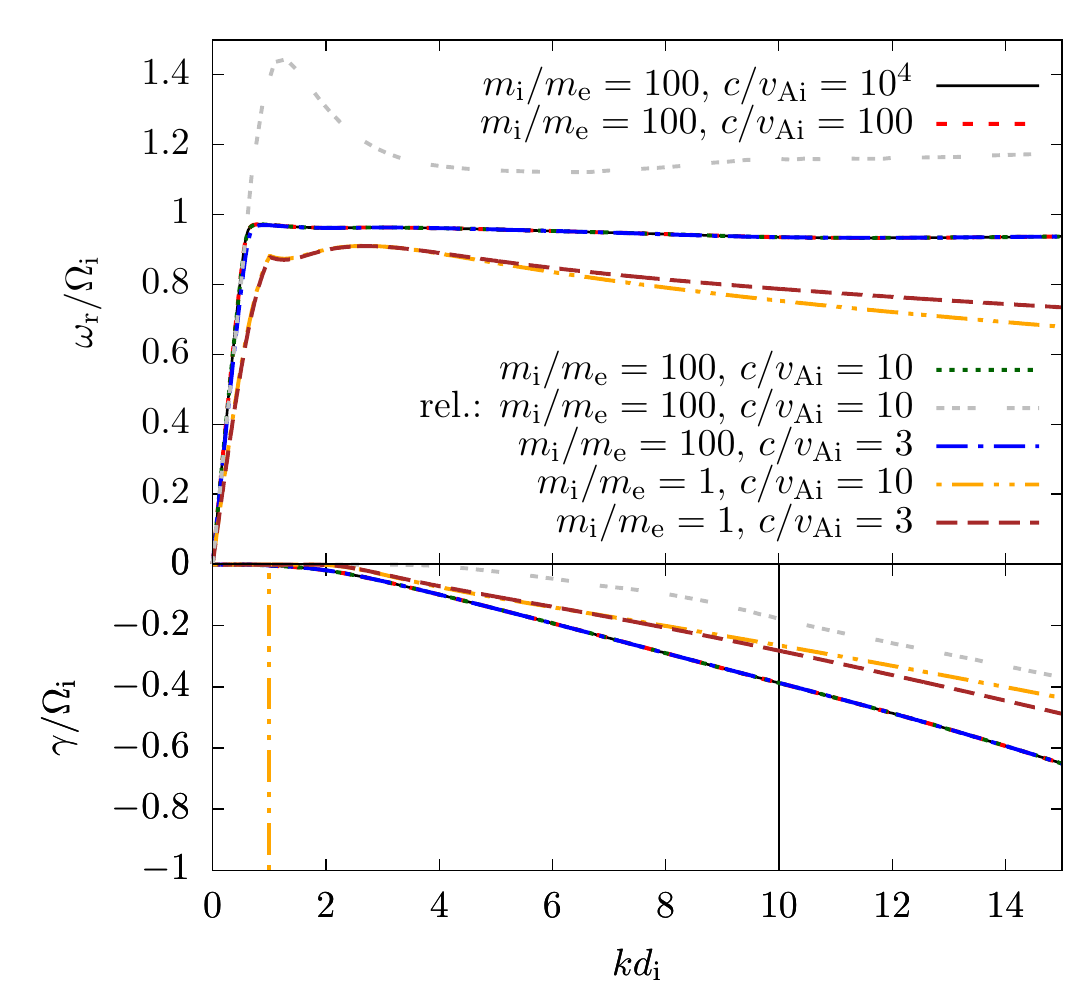}%
 \caption{Dispersion relations of the FM/W wave in quasi-perpendicular propagation for different values of $c/v_{\mathrm{Ai}}$.  We fix the ion-to-electron mass ratio at $m_{\mathrm i}/m_{\mathrm e}=100$ and $m_{\mathrm i}/m_{\mathrm e}=1$. The plot follows the same format as Fig.~\ref{perp_AIC_vA}. \label{perp_FMW_vA}}%
\end{figure}

In Fig.~\ref{perp_FMW_vA}, we show the dispersion relations for the FM/W wave in quasi-perpendicular propagation. We find that this mode's dispersion relation is largely independent of $c/v_{\mathrm{Ai}}$ for a constant value of $m_{\mathrm i}/m_{\mathrm e}$. The relativistic solution, however, shows a strong deviation from our non-relativistic solutions. For all $kd_{\mathrm i}$ shown in Fig.~\ref{perp_FMW_vA}, $|\gamma|/\Omega_{\mathrm i}$ is smaller than in the non-relativistic cases. However, $\omega_{\mathrm r}/\Omega_{\mathrm i}$ is greater in the relativistic case and exhibits a local maximum near $kd_{\mathrm i}\approx 1$. Although ion and electron scales are clearly separated in this case since $m_{\mathrm i}\gg m_{\mathrm e}$, relativistic electron effects modify the dispersion relation of the quasi-perpendicular FM/W wave at ion scales. 

\section{Discussion and Conclusions}

Using linear dispersion theory, we analyse the dependence of the normalized A/IC-wave and FM/W-wave dispersion relations in quasi-parallel and quasi-perpendicular propagation on $m_{\mathrm i}/m_{\mathrm e}$ and $c/v_{\mathrm{Ai}}$. In nonlinear computations such as particle-in-cell and Eulerian-Vlasov simulations, these two parameters are often artificially reduced compared to their actual values in nature. This reduction enables  simulations ranging from the outer scales of the system to the characteristic electron scales, which are reasonably smaller than the characteristic ion scales, while keeping computational costs small.

Using an ion normalization for frequencies and length scales, we find that the quasi-parallel A/IC wave is largely independent of $m_{\mathrm i}/m_{\mathrm e}$ and $c/v_{\mathrm{Ai}}$. In the quasi-perpendicular limit, $m_{\mathrm i}/m_{\mathrm e}$ has a stronger effect on both the wave frequency and the damping rate of the A/IC wave than $c/v_{\mathrm{Ai}}$. Relativistic effects also modify the A/IC-wave dispersion relation more strongly in the quasi-perpendicular limit than in the quasi-parallel limit.
Since the electron-cyclotron resonance defines the damping of quasi-parallel FM/W waves, their behaviour strongly depends on $m_{\mathrm i}/m_{\mathrm e}$, but less so on $c/v_{\mathrm {Ai}}$. Moreover, relativistic resonance effects increase the damping rate of the quasi-parallel FM/W wave.  In the quasi-perpendicular limit, the FM/W-wave dispersion relation is largely independent of $m_{\mathrm i}/m_{\mathrm e}$ over a wide range of $m_{\mathrm i}/m_{\mathrm e}$. Likewise, its dispersion relation does not exhibit a strong dependence on $c/v_{\mathrm{Ai}}$. We find, however, that relativistic electron effects have a strong impact on the dispersion relation of the quasi-perpendicular FM/W wave, even at ion scales.

The $c/v_{\mathrm{Ai}}$ dependence of all modes under consideration is small except for very small values of $c/v_{\mathrm{Ai}}\lesssim 3$. For these small values of $c/v_{\mathrm{Ai}}$, effects due to the displacement current modify the mode behaviour, especially for the quasi-parallel FM/W wave. We, therefore, recommend rather to aim for a more realistic mass ratio than for a more realistic value of $c/v_{\mathrm {Ai}}$ in kinetic plasma models if such a choice is necessary. There are two caveats to this conclusion: (i) relativistic effects, depending on the plasma parameters and the chosen normalization, gain importance with decreasing $c/v_{\mathrm {Ai}}$; and (ii) the scale separation between $d_{\mathrm i}$ and $\lambda_{\mathrm i}$ (likewise, depending on the plasma parameters and the chosen normalization, between $d_{\mathrm e}$ and $\lambda_{\mathrm e}$) decreases with decreasing $c/v_{\mathrm{Ai}}$, so that sub-Debye-length effects can be exaggerated at small scales when $c/v_{\mathrm{Ai}}$ is artificially small. Therefore, we extend our recommendation to include the consideration of unwanted relativistic and sub-Debye-length effects due to artificially low $c/v_{\mathrm {Ai}}$, see Eq.~(\ref{relgen}).


Our study suggests that plasma models with $m_{\mathrm i}/m_{\mathrm 
e}\gtrsim 100$ and $c/v_{\mathrm{Ai}}\gtrsim 10$ successfully cover 
physics on scales $\gtrsim 0.2d_{\mathrm i}$ for $\beta_{j}\sim 1$. 
Our results have two implications: (i) for astrophysical plasmas 
composed of electron--positron pairs, the magnetization does not 
significantly affect the linear behaviour; and (ii) in numerical simulations of 
ion--electron plasmas, an artificially smaller-than-realistic mass ratio $m_{\mathrm i}/m_{\mathrm e}$ affects the wave properties (frequency and damping rate) more strongly than an artificially smaller-than-realistic ratio $c/v_{\mathrm{Ai}}$. 

The various waves analysed here also depend on other plasma parameters such as $\beta_j$ of all species, the number of species, relative drifts among them, and their temperature anisotropies  \citep{marsch82,marsch82a,kasper02,bale09,maruca12,verscharen13a,yoon17,klein18}. It would be useful to repeat our investigation in individual cases when using plasma models with parameters different from the representative values used in this study  \citep{riquelme15,riquelme16,riquelme18}.

\section*{Acknowledgements}

The development of the \textsc{alps} code was supported by NASA grant NNX16AG81G. D.V.~is supported by the STFC Ernest Rutherford Fellowship ST/P003826/1 and STFC Consolidated Grant ST/S000240/1. T.N.P.~was supported by NSF SHINE grant AGS-1460130 during the completion of this work. S.P.G.~acknowledges support from NASA grant NNX17AH87G. K.G.K.~acknowledges support from NASA grant 80NSSC19K0912.

\bibliographystyle{mnras}
\bibliography{massratio_and_vAc_rev}

\bsp	
\label{lastpage}
\end{document}